\documentclass[preprint2]{aastex701}
\begin{document}

\title{The Critical 9365 {\AA} Diffuse Interstellar Band and C$_{60}^{+}$ Association}

\author[orcid=0000-0001-8803-3840]{Daniel Majaess}
\affiliation{Department of Chemistry and Physics, Mount Saint Vincent University, Halifax, Nova Scotia, B3M2J6 Canada.}
\email[show]{Daniel.Majaess@msvu.ca}

\author[orcid=0000-0003-3469-8980]{Tina A. Harriott}
\affiliation{Department of Mathematics and Statistics, Mount Saint Vincent University, Halifax, Nova Scotia, B3M2J6 Canada.}
\email{Tina.Harriott@msvu.ca}

\author{Halis Seuret}
\affiliation{Centro de Investigaciones Químicas, IICBA, Universidad Autónoma del Estado de Morelos, Cuernavaca, 62209, Morelos, Mexico.}
\email{halisseureth@gmail.com}

\author[orcid=0000-0002-8746-9076]{Cercis Morera-Boado}
\affiliation{IXM-Cátedra Conahcyt-Centro de Investigaciones Químicas, IICBA, Universidad Autónoma del Estado de Morelos, Cuernavaca, 62209, Morelos, Mexico.}
\email{cermor@gmail.com}

\author[orcid=0000-0001-6662-3428]{Lou Massa}
\affiliation{Hunter College \& the PhD Program of the Graduate Center, City University of New York, New York, USA.}
\email{lmassa@hunter.cuny.edu}

\author[orcid=0000-0001-8397-5353]{Ch\'erif F. Matta}
\affiliation{Department of Chemistry and Physics, Mount Saint Vincent University, Halifax, Nova Scotia, B3M2J6 Canada.}
\affiliation{Department of Chemistry, Saint Mary's University, Halifax, Nova Scotia, B3H3C3 Canada.}
\affiliation{D\'epartement de Chimie, Universit\'e Laval, Qu\'ebec, G1V0A6 Canada.}
\affiliation{Department of Chemistry, Dalhousie University, Halifax, Nova Scotia, B3H4J3 Canada.}
\email{Cherif.Matta@msvu.ca}

\begin{abstract}
The detection of interstellar C$_{60}^{+}$ has been debated for 30 years. The contested attribution of a weaker DIB at 9365 {\AA} was re-evaluated here on the basis of a Pearson correlation relative to 9577 {\AA}, which was previously tied to C$_{60}^{+}$ by diverse collaborations. An assessment of 11 sightlines revealed a high correlation amongst 9365$-$9577 {\AA} equivalent widths ($r=0.93 \pm 0.05$), after contamination from an adjacent line was mitigated using both numerical integration and Gaussian fits. In tandem with a recent separate study's high-$r$ evaluation linking 9577$-$9632 {\AA} across a sizable baseline: three interrelated DIBs matching C$_{60}^{+}$ laboratory findings were independently reaffirmed (9365, 9577, 9632 {\AA}).  Yet further investigations are required to strengthen the case via two other weak DIBs disputedly linked to C$_{60}^{+}$, particularly owing to potential overlapping lines arising from an expansive chemical space (PAHs).
\end{abstract}


\section{Introduction}
\citet{kr87} suggested C$_{60}^{+}$ could be associated with diffuse interstellar bands (DIBs).  Subsequent experimental and observational research indicated that 9577 and 9632 {\AA} DIBs may be tied to that fullerene \citep[e.g.,][]{cam15}.  However, concerns were expressed regarding the attribution \citep[e.g.,][]{gal21}.  Recently, a consensus is emerging that those two stronger DIBs are indeed highly correlated and stem from a common carrier \citep[e.g.,][]{nie22,maj25}.  Yet a convincing connection to the buckminsterfullerene cation remains tenuous when relying merely on two DIBs. Efforts are therefore ongoing to characterize weaker interstellar lines overlapping experimental C$_{60}^{+}$ measurements \citep[9348, 9365, 9428 {\AA}, e.g.,][]{ca16,co19}, but importantly those results are contested \citep{gk17,gal21}.

In this study, an independent assessment is undertaken regarding whether 9365 and 9577 {\AA} DIBs exhibit Pearson correlated equivalent widths (EW). The investigation is carried out by inspecting spectra presented by \citet{wa15,wa16}, \citet{gk17}, and \citet{co19}.  Specifically, the following sightlines are revisited: HD168625, HD136239, HD190603, BD$+$631964, 69 Cyg, HD169454, HD46711, Cyg OB2 5 (VI Cyg 5), Cyg OB2 12 (VI Cyg 12), HD183143, and HD195592 ($n=11$).
  
\section{Analysis}
\begin{figure}[t]
\begin{center}
 \includegraphics[width=3.4in]{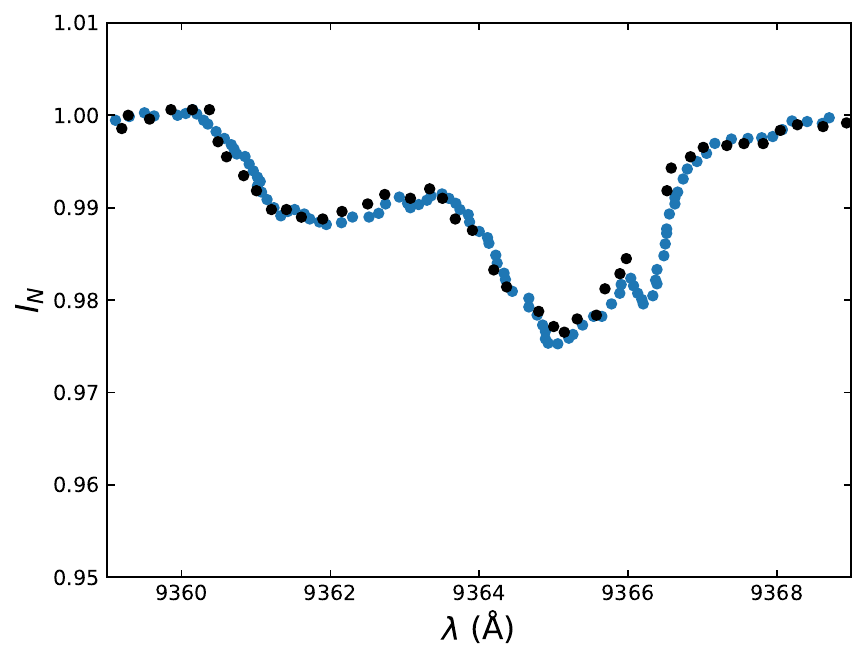} 
  \caption{The HD183143 sightline hosts the 9365 {\AA} DIB, and a proximate absorption line at 9362 {\AA}. Data were inferred from spectra displayed in \citet[][blue]{wa15} and \citet[][black dots, and omitting telluric features]{gk17}, and which represent an EW datum (average) in Fig.~\ref{fig2}. A minor telluric (de)contamination anomaly may exist near 9366.2 {\AA} in the former's spectrum.}
 \label{fig1}
\end{center}
\end{figure}

Wavelength and flux tables for the sightlines are absent from the assessed studies, hence diagrams therein were interpreted using PlotDigitizer. An example of the inferred data is shown in Fig.~\ref{fig1}, where generally consistent observations for HD183143 emerge when extracted from spectra displayed by \citet{wa15} and \citet{gk17}.  

Non-parametric numerical integration was favored for EW computations owing to spectral asymmetries and irregularities, which are exacerbated by molecular rotation. Simpson's approximation was implemented, and the 9365 {\AA} EWs were calculated near\footnote{$\lambda=9364.7$ {\AA}.} line center to longer wavelengths (i.e., $\approx \frac{1}{2}$EW), namely to reduce the impact of 9362 {\AA}. The example spectra in Fig.~\ref{fig1} show both 9362 and 9365 {\AA} absorption profiles, whereby the debated DIB linked to C$_{60}^{+}$ is 9365 {\AA}. The EW determination for 9577 {\AA} is comparatively uncomplicated relative to 9365 {\AA}. Simpson's approach was applied to the full 9577 {\AA} profile. A trapezoid approach was employed to help evaluate uncertainties, as discussed below, and validate the correlation determined together with Gaussian fits.

\begin{figure}[t]
\begin{center}
 \includegraphics[width=3.4in]{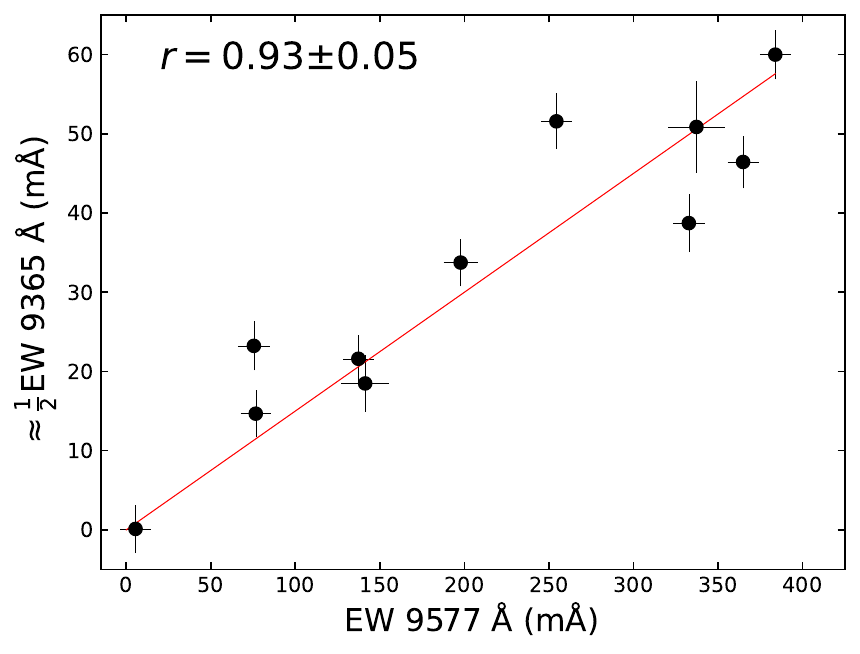} 
  \caption{The 9365 and 9577 {\AA} DIBs are highly correlated ($r=0.93\pm0.05$). The $\approx \frac{1}{2}$EWs for the 9365 {\AA} DIB represent numerical integration near line center to longer $\lambda$, thus mitigating contamination from the adjacent 9362 {\AA} (Fig.~\ref{fig1}). Importantly, a high correlation is confirmed by Gaussian fits ($r\simeq0.91$).}
 \label{fig2}
\end{center}
\end{figure}

HD169454, HD183143, and Cyg OB2 5 possess multiple spectra which provide a means of evaluating EW scatter. Data for HD169454 were presented by \citet{wa15}, \citet{gk17}, and \citet{co19}.  The 9365 {\AA} $\approx \frac{1}{2}$EWs determined here exhibit a standard deviation of 2 m{\AA}. For 9577 {\AA} EWs $\sigma =11$ m{\AA}.  Spectra for HD183143 were published by \citet{wa15} and \citet{gk17}, and $\sigma =2$ m{\AA} for both 9365 and 9577 {\AA}. Lastly, Cyg OB2 5 observations were published by \citet{gk17} and \citet{co19}, and the results are $\sigma=5$ and 14 m{\AA} for the DIBs, respectively. The average EW for each of those stars was utilized when constructing Fig.~\ref{fig2}. In an effort to estimate systematic uncertainties for the entire sample (stars with and without multiple spectra), a base bulk uncertainty was adopted for all stars from means of the aforementioned deviations (i.e., 3 and 9 m{\AA} for 9365 and 9577 {\AA} accordingly), and subsequently expanded in quadrature with any offset between Simpson and trapezoid integrations for each star. The uncertainties underestimate local effects, and for example note that HD195592 is reputedly a runaway star with bowshock induced $\gamma$-rays \citep[][]{val13}, which could explain its more deviant nature in Fig.~\ref{fig2} (EW$_{9577}= 254 \pm 9$ m{\AA}, and \citealt{co19} obtain $251\pm6$ m{\AA}), in concert with a normalization issue for 9365 {\AA}. Continuum normalization for HD195592 and BD$+$631964 are uncertain and exhibit potentially spurious shorter-wavelength shoulders \citep[e.g.,][their Fig.~2 and discussion therein]{co19}.  

In conclusion, 9365 and 9577 {\AA} are highly correlated (unweighted $r=0.93\pm0.05$, Fig.~\ref{fig2}). Pertinently, analytic Gaussian fits to 9365 {\AA} provided semi-independent confirmation of the high correlation ($r\simeq0.91$). Double Gaussians were employed to isolate 9362 {\AA} from the DIB 9365 {\AA}.  Moreover, the ensuing observed slope ($m_O=0.22$, EW$_{9365}$/EW$_{9577}$) is similar to indirect comparisons relative to the experimental attenuation ratio \citep[][$m_L=0.26$]{cm18}, thereby confirming that specific result of \citet{co19}.

Lastly, systematic uncertainties associated with 9362 {\AA} are currently too large to firmly evaluate whether it is correlated with 9365 {\AA} (Fig.~\ref{fig1}). For example, 9362 {\AA} profiles for Cyg OB2 5 differ between \citet{co19} and \citet{gk17}, and that is likewise true for HD169454 (\citealt{co19} relative to \citealt{wa15} and \citealt{gk17}). Utilizing all data yields $r=0.36\pm0.30$ (9362$-$9365 {\AA}), and excluding \citet{co19} spectra implies $r\approx 0.67$ (merely $n=5$). 

\section{Conclusions}
DIBs at 9365 and 9577 {\AA} feature EWs adhering to a high correlation ($r=0.93\pm0.05$, Fig.~\ref{fig2}), hence providing additional context relative to an existing and warranted broader C$_{60}^{+}$ debate. The result stems from non-parametric numerical integration and Gaussian fits, where contamination from a nearby absorption line was mitigated (Fig.~\ref{fig1}). Moreover, when paired with the \citet{maj25} high-$r$ determination between 9577$-$9632 {\AA} DIBs across a notable 650 m{\AA} baseline\footnote{Previous conflicting works each sampled a narrow dynamic range.}: three DIBs are now independently reaffirmed as interrelated (9365, 9577, 9632 {\AA}), and as noted previously likewise match experimental C$_{60}^{+}$ conclusions \citep{cam15,ca16}. 

Lastly, \citet{gal21} present pertinent arguments that evidence linking 9348 and 9428 {\AA} DIBs to C$_{60}^{+}$ is unconvincing, and this work did not address those two DIBs. Indeed, additional investigations of those two lines are desirable, since unknown PAHs could exhibit three comparable lines to C$_{60}^{+}$, but the probability diminishes substantially when considering five lines (i.e., wavelength and intensity ratios).  The current and envisioned analyses could be (in)validated and undertaken using telluric-corrected X-shooter spectra.

\bibliography{article}{}

\begin{thebibliography}{}
\expandafter\ifx\csname natexlab\endcsname\relax\def\natexlab#1{#1}\fi
\providecommand{\url}[1]{\href{#1}{#1}}
\providecommand{\dodoi}[1]{doi:~\href{http://doi.org/#1}{\nolinkurl{#1}}}
\providecommand{\doeprint}[1]{\href{http://ascl.net/#1}{\nolinkurl{http://ascl.net/#1}}}
\providecommand{\doarXiv}[1]{\href{https://arxiv.org/abs/#1}{\nolinkurl{https://arxiv.org/abs/#1}}}

\bibitem[{E.~K. {Campbell} {et~al.}(2015){Campbell}, {Holz}, {Gerlich}, \& {Maier}}]{cam15}
{Campbell}, E.~K., {Holz}, M., {Gerlich}, D., \& {Maier}, J.~P. 2015, \bibinfo{title}{{Laboratory confirmation of C$_{60}$$^{+}$ as the carrier of two diffuse interstellar bands},} \nat, 523, 322, \dodoi{10.1038/nature14566}

\bibitem[{E.~K. {Campbell} {et~al.}(2016){Campbell}, {Holz}, \& {Maier}}]{ca16}
{Campbell}, E.~K., {Holz}, M., \& {Maier}, J.~P. 2016, \bibinfo{title}{{C$_{60}$ $^{+}$ in Diffuse Clouds: Laboratory and Astronomical Comparison},} \apjl, 826, L4, \dodoi{10.3847/2041-8205/826/1/L4}

\bibitem[{E.~K. {Campbell} \& J.~P. {Maier}(2018){Campbell} \& {Maier}}]{cm18}
{Campbell}, E.~K., \& {Maier}, J.~P. 2018, \bibinfo{title}{{Isomeric and Isotopic Effects on the Electronic Spectrum of \{\{\textbackslashrm\{C\}\}\}\_\{60\}\^\{+\}-He: Consequences for Astronomical Observations of \{\{\textbackslashrm\{C\}\}\}\_\{60\}\^\{+\}},} \apj, 858, 36, \dodoi{10.3847/1538-4357/aab963}

\bibitem[{M.~A. {Cordiner} {et~al.}(2019){Cordiner}, {Linnartz}, {Cox}, {Cami}, {Najarro}, {Proffitt}, {Lallement}, {Ehrenfreund}, {Foing}, {Gull}, {Sarre}, \& {Charnley}}]{co19}
{Cordiner}, M.~A., {Linnartz}, H., {Cox}, N.~L.~J., {et~al.} 2019, \bibinfo{title}{{Confirming Interstellar C$_{60}$ $^{+}$ Using the Hubble Space Telescope},} \apjl, 875, L28, \dodoi{10.3847/2041-8213/ab14e5}

\bibitem[{M.~V. {del Valle} {et~al.}(2013){del Valle}, {Romero}, \& {De Becker}}]{val13}
{del Valle}, M.~V., {Romero}, G.~E., \& {De Becker}, M. 2013, \bibinfo{title}{{Is the bowshock of the runaway massive star HD 195592 a Fermi source?},} \aap, 550, A112, \dodoi{10.1051/0004-6361/201220112}

\bibitem[{G.~A. {Galazutdinov} \& J. {Kre{\l}owski}(2017){Galazutdinov} \& {Kre{\l}owski}}]{gk17}
{Galazutdinov}, G.~A., \& {Kre{\l}owski}, J. 2017, \bibinfo{title}{{Looking for the Weak Members of the $_{60}$$^{+}$C Family in the Interstellar Medium},} \actaa, 67, 159, \dodoi{10.32023/0001-5237/67.2.4}

\bibitem[{G.~A. {Galazutdinov} {et~al.}(2021){Galazutdinov}, {Valyavin}, {Ikhsanov}, \& {Kre{\l}owski}}]{gal21}
{Galazutdinov}, G.~A., {Valyavin}, G., {Ikhsanov}, N.~R., \& {Kre{\l}owski}, J. 2021, \bibinfo{title}{{Diffuse Bands 9577 and 9633: Relations to Other Interstellar Features},} \aj, 161, 127, \dodoi{10.3847/1538-3881/abd4e5}

\bibitem[{H.~W. Kroto(1987)Kroto}]{kr87}
Kroto, H.~W. 1987, Chains and Grains in Interstellar Space, ed. A.~L{\'e}ger, L.~d'Hendecourt, \& N.~Boccara (Dordrecht: Springer Netherlands), 197--206, \dodoi{10.1007/978-94-009-4776-4_17}

\bibitem[{D. {Majaess} {et~al.}(2025){Majaess}, {Harriott}, {Seuret}, {Morera-Boado}, {Massa}, \& {Matta}}]{maj25}
{Majaess}, D., {Harriott}, T.~A., {Seuret}, H., {et~al.} 2025, \bibinfo{title}{{Strengthening the link between fullerenes and a subset of diffuse interstellar bands},} \mnras, 538, 2392, \dodoi{10.1093/mnras/staf425}

\bibitem[{T.~P. {Nie} {et~al.}(2022){Nie}, {Xiang}, \& {Li}}]{nie22}
{Nie}, T.~P., {Xiang}, F.~Y., \& {Li}, A. 2022, \bibinfo{title}{{C$_{60}$ cation as the carrier of the {\ensuremath{\lambda}} 9577 {\r{A}} and {\ensuremath{\lambda}} 9632 {\r{A}} diffuse interstellar bands: further support from the VLT/X-Shooter spectra},} \mnras, 509, 4908, \dodoi{10.1093/mnras/stab3296}

\bibitem[{G.~A.~H. {Walker} {et~al.}(2015){Walker}, {Bohlender}, {Maier}, \& {Campbell}}]{wa15}
{Walker}, G.~A.~H., {Bohlender}, D.~A., {Maier}, J.~P., \& {Campbell}, E.~K. 2015, \bibinfo{title}{{Identification of More Interstellar C$_{60}$$^{+}$ Bands},} \apjl, 812, L8, \dodoi{10.1088/2041-8205/812/1/L8}

\bibitem[{G.~A.~H. {Walker} {et~al.}(2016){Walker}, {Campbell}, {Maier}, {Bohlender}, \& {Malo}}]{wa16}
{Walker}, G.~A.~H., {Campbell}, E.~K., {Maier}, J.~P., {Bohlender}, D., \& {Malo}, L. 2016, \bibinfo{title}{{Gas-phase Absorptions of C60+: A New Comparison with Astronomical Measurements},} \apj, 831, 130, \dodoi{10.3847/0004-637X/831/2/130}

\end{thebibliography}
\bibliographystyle{aasjournalv7}

\end{document}